# A padronização de frota e suas consequências para as companhias aéreas

Rodolfo Romboli Narcizo
Alessandro V. M. Oliveira⇥
Instituto Tecnológico de Aeronáutica, São José dos Campos, Brasil
⇥ Autor correspondente. Instituto Tecnológico de Aeronáutica. Praça Marechal Eduardo Gomes, 50. 12.280-250 - São José dos Campos, SP - Brasil.
E-mail: alessandro@ita.br.

*Resumo*: Este estudo aborda a crescente padronização das frotas aéreas, destacando que passageiros frequentes têm maior probabilidade de voar com o mesmo modelo de avião mais vezes. O objetivo é analisar a gestão de frotas das companhias aéreas e o impacto de uma variedade reduzida de modelos na operação das empresas. Discute-se os benefícios da padronização, como eficiência operacional, e os riscos, como vulnerabilidade a falhas específicas de um modelo. O trabalho revisa literatura científica internacional sobre o tema, identificando consensos e discordâncias que sugerem áreas para futura pesquisa. Inclui-se também um estudo sobre o mercado brasileiro, examinando como a padronização afeta custos operacionais e lucratividade nas dimensões de modelo, família e fabricante de aeronaves. Além disso, investiga-se a relação entre a padronização de frota e o modelo de negócios das companhias, concluindo que as vantagens da padronização não são exclusivas às empresas de baixo custo, mas também podem ser aproveitadas por outras companhias aéreas.

*Palavras-chave*: transporte aéreo, companhias aéreas, frota aérea.

## I. Introdução

Para o público consumidor de transporte aéreo em geral, o modelo de aeronave que irá realizar o voo talvez passe despercebido. Para os consumidores mais aficionados, entretanto, a emoção em realizar uma viagem aérea começa logo na compra da passagem, onde em muitos casos já é possível ter conhecimento dessa informação. A possibilidade de voar em uma nova aeronave e experienciar todo conforto e tecnologia embarcada são fatores de interesse para aqueles que apreciam, estudam ou trabalham no transporte aéreo em geral. Em alguns casos, chegam a ser inclusive fatores latentes muitas vezes determinantes para o nível de satisfação percebida pelo passageiro. Não por menos, aviões com histórico negativo tem sua denominação alterada para efeitos de marketing, como já pôde ser observada com o Fokker 100, rebatizado por MK-28 pela Avianca Brasil e mais recentemente o Boeing 737 MAX 8, rebatizado apenas por 737-8 pela Gol . Entretanto, muito antes que o modelo do avião possa ser indicado no website da companhia durante a materialização da venda da passagem, muitos aspectos foram bastante analisados e considerados pelo setor de planejamento estratégico da empresa aérea, de forma a garantir que a frota seja a mais eficiente e adequada ao plano de negócios e de malha.

A escolha de frota é ponto de suma importância, e tem impacto direto no custo da operação e no planejamento de malha de qualquer companhia aérea. Cada modelo de aeronave possui características próprias que serão determinantes para o sucesso da empresa, tais como o custo de aquisição ou arrendamento, custo de manutenção, custos de treinamento de tripulantes, consumo de combustível, alcance máximo de voo, capacidade de assentos e de carga, performance de decolagem e pouso, confiabilidade, conforto, entre tantos outros fatores.

As características operacionais de cada modelo de aeronave podem torná-lo mais adequado para um mercado específico de viagens aéreas, ao mesmo tempo em que também o torna ineficiente para outro. Tais pontos são considerados não somente no momento da decisão de compra de novas aeronaves, mas também durante a operação da empresa e no direcionamento da frota para cada destino. Em 2009 a companhia aérea britânica British Airways anunciou o lançamento de duas frequências diárias entre o aeroporto de Londres (London City), no Reino Unido, e o aeroporto de Nova Iorque (John F. Kennedy), nos Estados Unidos . Os voos, tiveram o encerramento anunciado em julho de 2020, durante a crise global provocada pelo vírus da COVID-19 , potencialmente em decorrência da queda brusca observada na demanda pelo transporte aéreo. O serviço premium era oferecido por aeronaves do modelo Airbus A318 configuradas com apenas 32 assentos de classe executiva totalmente reclináveis (em configuração de alta densidade, o modelo pode comportar até 132 assentos), garantindo um serviço exclusivo e com diversos benefícios para seus passageiros.

Além da quantidade limitada de passageiros, a rota também ganha destaque devido ao fato de que a distância de aproximadamente 5500 km entre Londres e Nova Iorque é muito superior aos 3700 km de alcance para um A318 em configuração máxima para transporte de passageiros/carga, conforme documento da Airbus . O voo somente era possível devido à configuração reduzida de passageiros e contava com uma escala para reabastecimento em Shannon, na Irlanda. Justamente devido à tais limitações, a colocação de aeronaves do porte de um A318 para realizar voos entre a Europa e a América do Norte não é comum. Rotas intercontinentais comumente são operadas por aeronaves com maior capacidade de assentos e maior autonomia, tal como os Airbus A330 e o Boeing 777, os quais podem transportar mais de 300 passageiros por distâncias superiores à 8000 km.

A escolha da empresa em operar uma rota de longa distância com uma aeronave de menor porte ao invés de aeronaves maiores, destaca a relevante relação entre a decisão de escolha de frota e o planejamento estratégico da companhia aérea. Entretanto, analisar as características individuais de cada modelo de aeronave, como alcance máximo, capacidade de transportar passageiros e cargas, consumo de combustível, entre outros, não é suficiente para escolher a composição de uma frota inteira, salvo se esta for composta por um único avião. A partir do momento em que há uma expansão e a quantidade de aeronaves aumenta, torna-se cada vez mais relevante para a companhia analisar o impacto que as aeronaves possuem quando consideradas em conjunto. Pode-se perceber que a quantidade de questionamentos que devem ser feitos começa a aumentar e o problema torna-se mais complexo, indo além da simples questão: "Qual modelo de aeronave mais adequado?" Neste ponto, a análise também deve ponderar: Quantos modelos optar? Modelos de um mesmo fabricante ou fabricantes diferentes? Famílias de aeronaves iguais, talvez? Qual o impacto de optar por mais de um modelo, família e fabricante? Como criar sinergia entre aeronaves de modelos ou fabricantes diferentes?

As opções para as companhias aéreas são diversas. Dentre os fabricantes mundiais de aeronaves comerciais, Boeing, Airbus, Embraer, Bombardier e ATR são os principais destaques. No portfólio de cada fabricante, é comum a presença de mais de uma família de aeronave, ação que tem por objetivo oferecer ao mercado aeronaves com perfis e características para diferentes tipos de missões. Disponibilizar diferentes famílias e modelos permite que um fabricante possa de atuar em diferentes áreas do mercado, desde empresas com rotas regionais até companhias com rotas intercontinentais. A Tabela 1 indica as principais famílias e modelos disponibilizados por cada fabricante.

**Tabela 1 – Principais fabricantes, famílias e modelos de aeronaves**

| Fabricantes | Famílias | Modelos | |
|---|---|---|---|
| Airbus | A320 | A318<br>A319<br>A320<br>A320 | A319neo<br>A320neo<br>A321neo |
| | A330 | A330-200<br>A330-300 | A330-800neo<br>A330-900neo |
| | A340 | A340-200<br>A340-300 | A340-500<br>A340-600 |
| | A350 | A350-800<br>A350-900 | A350-1000* |
| | A380 | A380-800 | |
| ATR | ATR42 | ATR42-500 | ATR42-600 |
| | ATR72 | ATR72-500 | ATR72-600 |
| Bombardier | C Series | CS100 (A220-100) | CS300 (A220-300) |
| | Q Series | Q400 | |
| Boeing | 737 | 737-600 NG<br>737-700 NG<br>737-800 NG<br>737-900 NG | 737-7 MAX<br>737-8 MAX<br>737-9 MAX<br>737-10 MAX |
| | 747 | 747-400 | 747-8 |
| | 767 | 767-200<br>767-300 | 767-400 |
| | 777 | 777-200<br>777-300 | 777-8*<br>777-9* |
| | 787 | 787-8<br>787-9 | 787-10 |
| Embraer | E170 | E170<br>E175 | E2 175 |
| | E190 | E190<br>E195 | E2 190<br>E2 195 |

*Modelo ainda em desenvolvimento.

Enquanto Boeing e Airbus disputam o mercado de aeronaves de médio e grande porte, Embraer e Bombardier visam o mercado de aeronaves de menor porte, para até 130 passageiros, com foco no intermédio entre o mercado regional e doméstico. A fabricante brasileira dispõe dos E-Jets da família E170 e dos modelos da família E190, os quais são maiores e possuem capacidade de acomodar mais passageiros e carga. Tais aeronaves foram desenvolvidas no final da década de 1990 e tiveram seus projetos atualizados para a construção dos E-Jets E2, jatos de nova geração com menor consumo de combustível e com redução de ruído. Em contrapartida, a canadense Bombardier possui os modelos da família C Series (posteriormente rebatizados pela Airbus), os quais também são projetos recentes e que empregam diversas tecnologias novas para redução do consumo de combustível. A fabricante também possui em seu portfólio o turboélice regional da família Q Series, que concorre diretamente com as aeronaves fabricadas pela ATR.

A europeia Airbus possui uma carteira ampla, com famílias voltadas para os mais diversos tipos de missões. Os quatro modelos da família A320 tem como foco voos de média distância e possuem entre 130 e 220 assentos. Tais aeronaves foram projetadas na década de 1980 e já possuem novas versões mais econômicas em operação, denominadas como Neo, em referência ao termo "New Engine Option". Para voos transoceânicos e de longa distância, a fabricante possui a família A330, família A340, família A350 e o modelo A380-800, o qual teve a produção descontinuada pela baixa demanda por parte das companhias aéreas.

Como concorrente direto à fabricante europeia, a Boeing também possui modelos e famílias para diferentes tipos de mercados. Para rotas de média distância com até 200 passageiros, a fabricante dos Estados Unidos dispõe da família 737. Os modelos Next Generation (NG) são reformulações da versão originalmente desenvolvida na década de 1960, e já foram aprimorados para a geração MAX. Para rotas de longa distância e maior densidade, a Boeing dispõe da família 747, família 767, família 777 e família 787.

Tanto a família 737, da Boeing, quando a família A320 da Airbus configuram entre as mais bem sucedidas do mundo. No ano de 2018, os modelos de médio alcance tiveram mais entregas do que qualquer outra família de aeronave, totalizando 579 aeronaves entregues pela Boeing e 626 entregas da fabricante europeia. Não obstante, as mesmas famílias também figuram entre as aeronaves com maior número de pedidos. No mesmo período, a família 737 somou 837 pedidos, enquanto a família A320 somou 577 pedidos.

Naturalmente, o desejo de qualquer empresa aérea é possuir uma frota que gere o maior retorno financeiro para a sua operação, sob o menor custo. Autores como Clark (2007), ex-gerente de treinamento e desenvolvimento da Airbus e mestre em Gerenciamento e Planejamento em Transporte pela University of Westminster, sugerem que a escolha e seleção de uma frota perfeitamente adaptável e flexível aos interesses da companhia aérea podem ser consideradas como um "Santo Graal", na qual o seu alcance beira quase o impossível. A razão para isso é de que o planejamento de frota compõe um problema multidimensional com diversas variáveis que estão em constante mudança, tanto por fatores de mercado quanto por condições macro e microeconômicas.

Neste trabalho, vamos abordar a padronização de frota e as consequências que ela pode ter para as companhias aéreas. Serão abordadas questões como a diferença conceitual entre comunalidade e padronização de frota; os efeitos que tais fatores podem ter em questões voltadas para o lado financeiro, como custos e geração de receita, para o lado operacional, como escala de tripulantes e malha aérea, e em aspectos estratégicos, como benefícios e potenciais riscos de ter uma frota padronizada; a padronização nas companhias aéreas brasileiras e a relação com o plano estratégico; os estudos internacionais de maior impacto no assunto; e os resultados de um estudo nacional

## II. LÓGICA DA PADRONIZAÇÃO

O termo comunalidade remete a ideia de comum. De forma geral, afirmar que uma frota contém um elevado índice de comunalidade é o mesmo que afirmar que essa mesma frota possui diversos fatores em comum. O termo é frequentemente utilizado para indicar a quantidade de peças e ferramentas equivalentes e intercambiáveis entre os seus próprios modelos, tal como para também indicar a similaridade operacional entre as aeronaves. Dado a relevância do tema para as companhias aéreas, o conceito de comunalidade é utilizado inclusive em portfólios e ações de marketing por parte dos fabricantes. A comunalidade entre as famílias E170 e E190 da Embraer chega a 89%, enquanto os jatos de uma mesma família podem alcançar até 95% . Segundo Alexandre de Figueiredo, vice presidente de operações da fabricante, um dos pontos de extrema importância para o desenvolvimento dos jatos E2 era justamente a comunalidade na operação com os jatos E1. Assim como a Embraer, os demais fabricantes também destacam a similaridades entre seus modelos. As aeronaves ATR42-600 e ATR72-600 possuem 90% de peças em comum . Os modelos CS100 e CS300 da Bombardier possuem o maior nível de comunalidade já conseguido por um fabricante, chegando a 99% de peças em comum . Os novos modelos das famílias A320 e A330 compartilham 95% das peças com os modelos de geração anterior. Por sua vez, as aeronaves 787 da Boeing foram projetadas para ter elevada comunalidade com as aeronaves da família 777, em especial na cabine de comando, com o objetivo de simplificar o treinamento dos tripulantes na transição entre os equipamentos .

Nota-se comunalidade é geralmente associada a aeronaves que pertençam a uma mesma família, já que estas foram desenvolvidas sob um mesmo projeto-base e tendem a possuir maior compartilhamento do que aeronaves de projetos distintos. Entretanto, ressalta-se a existência de uma ligeira diferença conceitual entre os termos comunalidade e padronização de frota. Enquanto o primeiro tende a indicar e quantificar as similaridades de peças e componentes entre duas ou mais aeronaves, o segundo termo apenas remete à ideia de busca por um padrão único, o qual pode ser analisado e quantificado de diversas maneiras, dependendo apenas de qual padrão está sendo buscado.

A diferença conceitual entre ambos os termos já foi brevemente abordada por pesquisadores da literatura, tal como Brüggen e Klose (2010) da Maastricht University, e tem por base a hipótese de que uma frota pode ser completamente padronizada enquanto ainda assim pode não estar no mais elevado índice de comunalidade possível. Imagine uma frota hipotética inteiramente composta por aeronaves do modelo Airbus A320. Isso faz com que esta seja inteiramente padronizada em modelo de aeronave. Entretanto, para este modelo, a fabricante Airbus autoriza tanto o emprego dos motores CFM56, da CFM International, quanto os IAE V2500, da International Aero Engines.  Visto que os modelos dos motores são distintos e de fabricantes diferentes, essa mesma frota pode ter índices de comunalidade diversos dependendo de como foi motorizada. Por outro lado, imagine agora uma frota composta por aeronaves dos modelos Boeing 737 e Airbus A320. Ambos podem ser equipados com motores CFM56. Isso permite, embora de forma pequena, que a comunalidade de peças e ferramentas ainda exista, muito embora a padronização de frota seja mais baixa.

Os exemplos com os motores tratam de um item de custo elevado e com impacto expressivo na operacionalidade da companhia. Entretanto, note que a ausência ou presença de comunalidade pode constar nos mais variados sistemas da aeronave, tal como nos winglets, sistema de entretenimento, configuração interna, autonomia de voo e especificações da aeronave. Esses e outros tantos pontos podem variar de forma significativa dentro de uma mesma frota, fazendo com que, quanto mais aspectos divergentes e peças não-intercambiáveis existirem entre as aeronaves, menor será a presença da comunalidade na frota. Essencialmente, a individualidade de cada aeronave não alterará a padronização da frota enquanto os modelos de aeronaves desta forem os mesmos. Por outro lado, manter uma padronização não garante que as aeronaves sejam totalmente comuns.

Mensurar a comunalidade em sua totalidade exige tamanha demanda por estatísticas e informações que acaba tornando inviável sua utilização em pesquisas. Os pontos potencialmente distintos entre aeronaves são diversos, fazendo com que esta seja uma variável latente e difícil de ser inteiramente observada. Como senso comum, autores propõem a captura dos aspectos da comunalidade através da própria padronização de frota. A suposição para isso é justamente a de que aeronaves de um mesmo fabricante provavelmente terão maior comunalidade de peças do que aeronaves de fabricantes diferentes. Esta probabilidade seria naturalmente intensificada conforme a padronização ficasse ainda mais restrita e passasse para o nível de família e, posteriormente, de modelo de aeronave. Apesar da diferença conceitual entre os termos, ambos são comumente utilizados como sinônimos por pesquisadores para representar os efeitos e consequências que uma frota padronizada tem para uma companhia aérea.

Há uma frequente associação na literatura acadêmica entre a padronização de frota e a redução de gastos e aumento na eficiência operacional. Manter uma frota padronizada pode reduzir gastos com tripulantes, despesas com manutenção e aumentar a capacidade da empresa em caso de necessidade de recuperação de escala (BOX 2) (Holloway, 2008; Berrittella; La Franca; Zito, 2009; Zou; Yu; Dresner, 2015). Vide, por exemplo, que cada modelo de aeronave exige que os tripulantes tenham treinamentos e habilitações específicas. Isso faz com que o número de funcionários necessários para operar uma aeronave seja diretamente influenciado pela quantidade de modelos de aeronaves diferentes de uma frota. Manter aeronaves iguais na frota garante similaridade na operação e permite um maior compartilhamento de pilotos e comissários. Tal ponto reduz os custos com treinamentos, têm o potencial de aumentar a produtividade da empresa e reduz a complexidade de gerenciamento da escala de voo dos tripulantes.

No que diz respeito às despesas de manutenção, cada aeronave possui componentes críticos que precisam estar em correto funcionamento para que o voo seja efetuado. Para reduzir as chances de cancelamentos, as companhias aéreas frequentemente possuem em estoque nas suas bases operacionais as principais peças e ferramentas da aeronave. A padronização aumenta a chance de comunalidade e permite que um maior número de aeronaves compartilhe de um mesmo estoque, reduzindo a quantidade de peças de reposição por aeronave da frota. Adicionalmente, uma frota com elevada comunalidade proporciona, além de componentes iguais, similaridade nos sistemas das aeronaves, o que pode aumentar a familiaridade dos mecânicos e engenheiros com os procedimentos de manutenção necessários e possibilita que desempenhem sua função com maior eficiência.

Por sua vez, o aumento na capacidade de recuperação de escala pode ser notado em situações onde ocorram atrasos e cancelamentos de voos. Em cenários como esse, a companhia aérea fica obrigada a rearranjar suas aeronaves de forma a reduzir os prejuízos em sua malha aérea. Neste caso, a presença de uma frota padronizada, ou seja, com características comuns - como, por exemplo, mesma quantidade de assentos ou autonomia de voo - pode simplificar o processo de recuperar a escala de voo após a retirada de uma aeronave de serviço.

Além destes supostos benefícios, a padronização de frota é comumente associada às companhias aéreas de baixo custo. Diversas companhias LCC nos mais variados continentes adotam a política de padronização e operam com modelos de aeronaves de uma única família, tal como a pioneira Southwest Airlines, dos Estados Unidos, a qual é um dos maiores clientes da Boeing, e o maior comprador de aeronaves da família 737 da história, tendo adquirido mais de 800 aviões desde 1971 . Outras empresas, como irlandesa Ryanair, optam por ser ainda mais restritivas e utilizam-se apenas um único modelo de aeronave. A companhia já recebeu mais de 500 aeronaves do modelo 737-800 desde 1999 e é o terceiro maior comprador da família. Note que, no exemplo da empresa europeia, apesar da tentativa de aumentar a sinergia de custos, a companhia fica restrita a uma frota de aeronaves com um mesmo alcance, quantidade de assentos e capacidade de carga. Isso faz com que haja uma maior limitação da empresa em responder a ações concorrenciais e a variações na demanda. Manter uma padronização excessivamente elevada tem potencial para reduzir a flexibilidade da companhia em responder rapidamente ao mercado, assim como também restringe a quantidade de rotas financeiramente rentáveis para àquelas adequadas a aeronaves com alcance e tamanho específico.

À parte de aspectos operacionais e assim como qualquer processo de compra e venda, a padronização também tem potencial para influenciar acordos entre companhia aérea e fabricante de aeronave. Pontos como valor de aquisição, formas de financiamento, tempo de espera até a entrega, desconto sobre grandes encomendas e tantos outros estão diretamente ligados à capacidade de negociação dos representantes de cada uma das empresas. Se por um lado manter a padronização de frota destaca a importância de o fabricante reconhecer a fidelização de um cliente e valorizá-lo com preços competitivos, por outro, adotar a padronização sugere certa inércia da companhia aérea em querer substituir essa padronização e uma relutância em alterar seus fornecedores, visto que esta modificação pode impor à empresa grandes esforços em aprender sobre as novas tecnologias empregadas nas novas aeronaves e em compreender a cultura e a forma de trabalho do novo fornecedor (Clark, 2007).

Dessa forma, o poder de barganha de cada um dos envolvidos pode ser a diferença entre conseguir apenas um acordo bom, e conquistar uma compra excelente, onde a empresa sai, não somente com um bom negócio, mas também com vantagens significativas perante seus concorrentes. Segundo Scott Kirby, presidente da American Airlines entre dezembro de 2013 e agosto de 2016, a companhia obteve, em acordo com a Airbus, o que o próprio executivo chamou de "cláusula de cliente mais favorecido" . No pedido de 260 aeronaves da família A320 que a empresa realizou em 2011, o fabricante comprometeu-se a reembolsar a companhia aérea com a diferença de valor caso uma mesma aeronave fosse vendida a outra empresa por um preço inferior.

Tal acordo apenas enaltece a capacidade da companhia aérea de barganhar pelo produto desejado e coloca até mesmo o fabricante de aeronaves em uma situação possivelmente delicada perante as outras empresas aéreas. Independentemente de ser ou não consequência deste acordo, 78% das encomendas feitas pela United (empresa concorrente da American Airlines) à época eram de aeronaves da Boeing. Em 2016, logo após sair da American Airlines, Scott Kirby assumiu o cargo de presidente da United e afirmou que a fabricante europeia Airbus não conseguia fornecer aeronaves à United por um preço competitivo, justamente devido ao acordo mencionado anteriormente.

Além de ponderar sobre os benefícios operacionais e a possibilidade de elevação do poder de barganha da companhia aérea, a decisão sobre a padronização de frota deve ser tomada com cautela e considerar potenciais riscos para a empresa. Em março de 2019, a frota mundial do recém-lançado Boeing 737 Max, ficou impossibilitada de voar devido a dois acidentes com vítimas fatais em curtos intervalos de tempo . Mais de ano depois, em agosto de 2020, o modelo ainda estava proibido de efetuar voos comerciais, e aguardava a emissão de uma nova certificação para voos. O impacto financeiro para o fabricante foi calculado em mais de 25 bilhões de dólares , incluindo 5,6 bilhões somente em compensação para companhias aéreas. As renovações e incrementos de frotas previstas para acontecer com as entregas dos modelos MAX de nova geração foram frustradas e tiveram impacto direto na operação de companhias aéreas do mundo todo, acarretando no cancelamento de milhares de voos programados para ocorrer no período.

Empresas como Southwest Airlines, nos Estados Unidos, Ryanair, na Irlanda, Flydubai, nos Emirados Árabes Unidos e Gol, no Brasil, operam com frotas exclusivas de aeronaves da família 737. Isso faz com que, devido à falta de variabilidade de aeronaves, estas sofram mais os impactos causados pela proibição. No caso da companhia brasileira, a concorrente Avianca Brasil havia encerrado suas operações em maio de 2019 , pouco tempo depois da proibição de voos do 737 MAX. O mercado em potencial deixado pela empresa tornou-se um desafio para a Gol, a qual, por não poder contar com os modelos de nova geração que havia encomendado, teve que tomar a decisão de arrendar aeronaves mais antigas no mercado secundário .

Justamente com o objetivo de evitar problemas como este, algumas companhias aéreas optam por reduzir a comunalidade e padronização de frota de forma a diminuir a possibilidade de proibição de voo de todas as suas aeronaves. Em fevereiro de 2020, a japonesa All Nippon Airlines anunciou a aquisição de quinze novas unidades do Boeing 787 Dreamliner. Entretanto, em sua nova encomenda a empresa optou por motores General Eletric (GE), diferentemente das aeronaves já em operação na empresa que operam com motores Rolls-Royce . Isso aumenta para a companhia aérea a segurança de que a frota não ficará inteiramente no solo em caso de problema com os motores ou fabricantes. No Brasil, a Latam é exemplo de situação similar, visto que em sua frota há aeronaves A320ceo tanto com motores V2500, produzidos pela International Aero Engines AG, quanto com motores CFM56, da CFM International.

*A padronização de frota nas companhias aéreas brasileiras*

A partir do encerramento da operação da Avianca Brasil, em 2019, o Brasil passou a contar com três companhias aéreas nacionais para suprir a necessidade por voos domésticos. Em uma análise não tão complexa e avançada da frota dessas empresas, já é possível notar diferentes estratégias de negócio e de malha. As companhias Azul, Gol e Latam iniciaram suas operações em 2020 com as seguintes composições de frota:

**Tabela 2 – Frota das companhias brasileiras em janeiro de 2020**

| Companhia | Modelo | Assentos | Total | Variabilidade | | | Idade média (em anos) |
|---|---|---|---|---|---|---|---|
| | | | | Modelos | Famílias | Fabricantes | |
| Azul | 1 A320CEO | 174 | 143 | 10 | 5 | 4 | 6,0 |
| | 38 A320NEO | 174 | | | | | |
| | 1 A321NEO | 214 | | | | | |
| | 8 A330-200 | 271 | | | | | |
| | 2 A330-900NEO | 298 | | | | | |
| | 31 ATR72-600 | 70 | | | | | |
| | 6 E190 | 106 | | | | | |
| | 51 E195 | 118 | | | | | |
| | 3 E195 E2 | 136 | | | | | |
| | 2 B737-400F | Carga | | | | | |
| Gol | 23 B737-700 | 138 | 138 | 3 | 1 | 1 | 9,8 |
| | 108 B737-800 | 186 | | | | | |
| | 7 B737-8 MAX | 186 | | | | | |
| Latam | 22 A319 | 144 | 160 | 7 | 4 | 2 | 9,0 |
| | 70 A320 | 174 | | | | | |
| | 6 A320NEO | 174 | | | | | |
| | 31 A321 | 220 | | | | | |
| | 8 A350 | 339 | | | | | |
| | 13 B767-300 | 238 | | | | | |
| | 10 B777-300 | 379 | | | | | |

A Azul, companhia considerada como LCC pela ICAO e nas categorias de premiação da Skytrax, possui a frota mais diversificada dentre as empresas. Ao todo são 10 modelos de aeronaves diferentes, na qual 6 são de gerações diferentes. Essa variabilidade de modelos representa também elevada quantidade de famílias disponíveis para a operação da empresa. As 5 famílias integrantes da frota têm características bastante distintas e permitem que diferentes mercados possam ser operados pela companhia de forma eficiente. A aeronaves ATR tem capacidade para até 78 passageiros e permite que a empresa tenha uma malha regional e opere em destinos com menor demanda ou em aeroportos com menor infraestrutura. Os jatos Embraer da empresa operam entre 118 e 136 passageiros, e também são mundialmente reconhecidos e utilizados para a realização de voos regionais onde a demanda e capacidade da rota já permitem a operação de aeronaves superiores aos turboélices, porém ainda inadequadas para aeronaves maiores, como as da família A320, configurados com capacidade entre 174 e 214 passageiros. Já os wide-bodies (BOX 1) da família A330 tem como foco voos internacionais, dada a sua autonomia e capacidade para até 298 passageiros.

É sabido que a oferta deve acompanhar a demanda, e a elevada variabilidade da frota permite com que a empresa adeque sua oferta para a demanda específica de uma rota. Dessa forma, abre-se a possibilidade para que a operação chegue a um maior número de destinos de características variadas, desde rotas regionais a partir de cidades menores como Aracati, CE e Tefé, AM, passando por grandes centros urbanos como Recife, PE e Brasília, DF, até destinos internacionais de longa distância, como Lisboa, em Portugal e Orlando, nos Estados Unidos. A seleção de destinos e rotas a serem operadas são também são decisões complexas e levam em consideração diversos fatores. O tópico será mais bem abordado no Trabalho "Por que a minha cidade não tem voos? Determinantes do planejamento estratégico da malha aérea de uma empresa". Ressalta-se neste momento, apenas que a extensa possibilidade de malha para a Azul tem grande responsabilidade da frota operada pela companhia.

Com uma frota bem mais homogênea, a também considerada low-cost Gol opera com apenas 3 modelos de uma única família de aeronaves da Boeing. A diferença entre as tratativas de ambas as companhias para a escolha da frota, sugere diferenças de modelo de negócios e no plano estratégico. A baixa variabilidade de modelos da Gol faz com que a empresa tenha por objetivo simplificar a operação, através da redução de ferramentas e materiais necessários, diferentes treinamentos para tripulantes e facilidade de readaptar a malha em caso de ruptura. A escolha da frota também sugere a diferença de mercado em que Azul e Gol operam. Enquanto a primeira tem grandes ramificações em mercado regionais, a Gol tem como foco rotas de maior densidade, visto que a maior parte das suas aeronaves contam com 186 assentos. A exceção fica por algumas rotas que podem ser operadas pela empresa com uma menor versão da família 737, a qual é configurada com 138 assentos.

A maior frota do país pertence a companhia Latam. A empresa não dispõe de aeronaves para realização de voos regionais., porém também opera com maior variabilidade de modelos e famílias. A maior parte das suas operações é, assim como a Gol, focada em mercados de maior densidade, considerando que suas principais aeronaves operam com 174 até 220 assentos. A menor aeronave que a empresa possui faz parte da família A320 e é configurada para 144 assentos, o que a impede de operar em aeroportos regionais e com menor demanda, tal como a estratégia adotada pela Azul. Por outro lado, a maior quantidade de aeronaves de grande porte e adequadas para a execução de voos de longa distância indica uma vertente estratégica também direcionada para voos internacionais. A Latam é a companhia brasileira com maior quantidade de rotas para cidades na Europa e América do Norte, as quais seriam inviáveis, por razões óbvias, em uma frota que priorizasse aeronaves de porte regional.

É notável a diferença de frota e de estratégia de negócios adotada por cada empresa. Enquanto a padronização é fielmente utilizada por uma, a diversidade de frota é ponto chave para a malha e plano estratégico de mercado para outras. Apesar disso, um ponto em comum entre as três empresas é a substituição de modelos mais antigos da frota por aeronaves de geração mais nova. Segundo os planos divulgados pela Azul, todos os jatos Embraer devem ser substituídos até 2021 pelas mais modernas e mais eficientes aeronaves de segunda geração, conhecidos como E2. A Gol possui encomenda de ao menos 135 Boeing 737 Max para substituição dos 737-800 NG, enquanto que a Latam gradualmente tem substituído as aeronaves da família A320ceo por modelos Neo de nova geração.

A renovação das frotas é movimento natural que tem por objetivo reduzir custos operacionais e incrementar a oferta de assentos e de carga, visto que os modelos mais modernos possuem menor consumo de combustível, menor produção de ruído, maior capacidade de carga e melhor aproveitamento de áreas internas, o que eleva a quantidade de assentos. A substituição de aeronaves por modelos iguais de nova geração traz a facilidade de que as aeronaves ainda contam com similaridades na operação e nos materiais e ferramentas utilizados, o que acaba por reduzir os gastos com treinamentos de adaptação de tripulantes, mecânicos e funcionários de solo. Entretanto, apesar da elevada comunalidade entre modelos iguais de gerações diferentes, ainda é necessário para as empresas a aquisição de peças e ferramentas não comuns entre as aeronaves com o objetivo de adequar suas bases operacionais para a entrada em serviço dos novos modelos. Tais adaptações de materiais e de funcionários também representam custos para a companhias, os quais devem ser considerados na decisão de renovar a frota.

*O que dizem os estudos da área*

A padronização de frota é tópico que já foi abordado na literatura sob diversas vertentes. A própria relação entre padronização e o modelo de negócio low cost é ideia bastante difundida e comumente aceita na literatura mundial (Gillen, 2006; Kilpi, 2007; Berrittella; La Franca; Zito, 2009; Brüggen; Klose, 2010; Zuidberg, 2014; Zou; Yu; Dresner, 2015; Daft; Albers, 2015). Autores como Kilpi (2007), da Finnair, e Brüggen e Klose (2010), da Maastricht University, afirmam, através de evidências analíticas, que a padronização de frota é comumente alta entre companhias LCC. Entretanto, em seu estudo, Kilpi (2007) faz a ressalva de que a padronização das companhias aéreas diminuiu ao longo dos últimos 50 anos, ao passo que o tamanho médio das frotas aumentou, representando um maior interesse das empresas em manter frotas com mais modelos de aeronaves.

Um dos trabalhos pioneiros sobre efeitos da padronização nos custos é de Seristö e Vepsäläinen (1997), de Helsinki School of Economics and Business Administration. Os autores questionam se a padronização de frota de fato traz benefícios às companhias aéreas e, ao analisarem a composição de frota de 42 empresas, encontram um efeito negativo entre a padronização de frota e os custos de voo e de manutenção de aeronaves. Entretanto, vale ressaltar que os efeitos sob os custos de manutenção foram considerados muito pequenos, o que seria um indicativo de que a real economia da padronização de frota seria proveniente dos custos de voo.

Similarmente, Zuidberg (2014), SEO Economic Research, também encontra em sua análise econométrica um efeito pequeno para manutenção. A autor obtém resultados que indicam um efeito negativo entre custos operacionais e padronização de frota. A princípio, isso poderia indicar que, quanto maior fosse a padronização da companhia aérea, menores seriam os custos da empresa. O autor ressalta que os resultados apresentam pouca significância estatística, levando-o a concluir que poucas evidências haviam sido encontradas que pudessem indicar algum efeito da padronização de frota sobre os custos operacionais.

O estudo internacional mais recente e de maior impacto no assunto, é de Zou, Yu e Dresner (2015). Os dois autores da Embry-Riddle Aeronautical University e o pesquisador da University of Maryland, respectivamente, analisam os efeitos da padronização de frota de forma segmentada. A análise é executada de forma a identificar o impacto de cada um dos níveis de padronização de frota: fabricante, família e modelo de aeronave. Os autores analisam por meio da econometria dados de 28 companhias aéreas dos EUA no período de 1999 a 2009. Segundo os resultados obtidos, frotas padronizadas em nível de família e/ou de modelo apresentariam menores custos unitários, ao passo que a padronização por fabricante acabaria por não ter efeitos.

Ainda no mesmo estudo, Zou, Yu e Dresner (2015) também investigam os efeitos da padronização na margem de lucro das companhias aéreas e encontram uma correlação positiva quando a empresa possui padronização de família. A hipótese para tal achado é de que uma família de aeronave forneceria à companhia aérea a flexibilidade necessária para se adaptar em situações de variação na demanda e de ações concorrenciais, sem que a frota perdesse em comunalidade. Isso seria possível graças ao fato de que as aeronaves de uma mesma família apresentam características em comum enquanto também possuem capacidades de assentos diferentes. Em outras palavras, segundo o estudo, a padronização de frota, quando adequadamente aplicada, teria a capacidade tanto de aumentar o potencial de obtenção de receita quanto o de reduzir os custos operacionais, elevando, portanto, a margem de lucro da companhia aérea. Kilpi (2007) também afirma que companhias aéreas que possuem frota mais padronizada teriam a tendência de apresentar melhores resultados financeiros, justamente devido a uma elevada correlação entre padronização e lucro operacional. Entretanto, diferentemente de Zou, Yu e Dresner (2015), o autor não explica se o aumento no lucro observado seria resultante de uma redução nos custos operacionais ou se devido a um aumento na receita da empresa.

Resultados similares foram encontrados no estudo de West e Bradley (2008), da East Carolina University. Segundo os autores, uma frota menos padronizada seria vantajosa para a companhia aérea quando esta tivesse o interesse de manter aeronaves de diferentes características para poder operar em mercados com demandas distintas, mas ainda assim possuir uma taxa de ocupação elevada. A princípio, tal estratégia teria o potencial de elevar o lucro da empresa e assemelha-se com as ideias de flexibilidade apresentadas por Zou, Yu e Dresner (2015). Entretanto, como os resultados obtidos por West e Bradley (2008) indicaram maiores lucros quando a padronização é maior, os autores sugerem que os custos decorrentes de possuir muitos modelos de aeronave sobrepõem os benefícios alcançados ao combinar melhor a oferta e demanda. No estudo, as evidências obtidas sugerem que a padronização de frota está positivamente correlacionada com margem de lucro.

Além da análise sobre a relação com modelo de negócios, Brüggen e Klose (2010) também estudam qual o efeito da padronização de frota no desempenho operacional de companhias aéreas. Para este tópico, os autores indicam que a padronização de frota está positivamente correlacionada com o desempenho operacional, e afirmam ainda que o tamanho da frota seria um fator intensificador dessa relação, fazendo com que, quanto maior a frota, melhor seria o desempenho da empresa. A justificativa dada para tais resultados seria de que frotas padronizadas poderiam aprimorar os trabalhos técnicos feitos na companhia aérea e que o tamanho traria consequências através do ganho de escala. Merkert e Hensher (2011), da Cranfield University e The University of Sidney, respectivamente, afirmam que a padronização de frota tem a capacidade de aumentar a eficiência operacional das companhias aéreas. Os autores observaram uma correlação negativa entre o número de famílias presentes em uma frota e a eficiência técnica, de alocação de recursos e de custos das empresas. O mesmo resultado foi obtido com relação ao número de fabricantes de aeronave. De forma similar, Weiss e Maher (2009), da Tel Aviv University e University of California, respectivamente, sugerem que a padronização de frota eleva o nível de cobertura das companhias aéreas para fornecer uma melhor capacidade de resposta para a ocorrência de eventos inesperados.

Boas, Cameron e Crawley (2012) de Massachusetts Institute of Technology (MIT), discute se os possíveis benefícios da padronização permanecem iguais durante a vida útil de um produto. A ideia por trás disso é de que ao longo do tempo, cada aeronave passaria por diferentes situações, teria operações distintas e poderia sofrer diversas adaptações. Mesmo os procedimentos de reparo e manutenção determinados pelo fabricante poderiam ter especificidades àquele modelo, o que poderia, ao longo da vida útil da aeronave, reduzir o nível de comunalidade e, consequentemente, seus potenciais benefícios. Tal fenômeno é denominado pelos autores como divergência da comunalidade. Adicionalmente, Boas, Cameron e Crawley (2012) também questionam o impacto de um deslocamento temporal na produção de dois equipamentos iguais. Segundo os autores, por mais que duas aeronaves sejam de um mesmo modelo, a produção destas em momentos diferentes no tempo reduziria o comunalidade existente entre as duas. Isso poderia ocorrer, uma vez que o próprio projeto da aeronave poderia sofrer modificações e incrementos, assim como o processo de produção poderia ser alterado.

Um dos pontos que possuem diferentes abordagens entre os autores é a maneira de tratar e calcular a padronização de frota. Borges Pan e Espírito Santo Junior (2004), da Universidade Federal do Rio de Janeiro, segmentam-na de acordo com partes das aeronaves, como célula e motor, e estabelecem índices e valores de ponderação para o efeito de cada parte na comunalidade. Zuidberg (2014) pondera a comunalidade através da quantidade de modelos de aeronaves na frota, número total de aeronaves e a quantidade de aeronaves do modelo mais comum. O estudo mais recente é baseado no Herfindahl-Hirshman Index (HHI), índice comumente utilizado para cálculo da concentração de mercado. O método proposto por Zou, Yu e Dresner (2015), sugere a segmentação da padronização em níveis de fabricante, família e modelo de aeronave, onde os resultados variam de 0 a 1 e, quanto mais próximo de zero, menor é a padronização de frota observada. Como cada fabricante possui em seu portfólio diversas famílias e, em cada uma destas há diferentes modelos de aeronaves, podemos dizer que uma frota hipotética que apresenta elevada padronização por modelo de aeronave também possuirá, por consequência, elevada padronização de família e fabricante. Note, entretanto, que o processo inverso não ocorre e uma padronização elevada por fabricante não indicará necessariamente uma elevada padronização por família e tampouco por modelo de aeronave..

### III. Estudos Brasileiros

No Brasil, o Núcleo de Economia do Transporte Aéreo do Instituto Tecnológico de Aeronáutica (NECTAR-ITA) tem realizado estudos em que aspectos da padronização de frota são abordados. A relação entre padronização de modelo de aeronaves e o número de destinos servidos, por exemplo, é abordada por Narcizo e Oliveira (2018) e Narcizo, Oliveira e Dresner (2020). A correlação negativa encontrada no estudo entre tais fatores indica que quanto maior a malha da companhia, maior a necessidade de adequar as aeronaves aos mercados e distância das rotas operadas, indicando menor padronização de modelo de aeronave. Os autores sugerem também que frotas mais velhas tendem a apresentar menor padronização de modelo, decorrente da busca ou substituição por modelos mais novos e que sejam mais eficientes e tecnologicamente mais avançados. O mesmo estudo contrapõe o resultado de Kilpi (2007) de que a padronização tem reduzido ao longo dos últimos 50 anos. Nos resultados divulgados por Narcizo e Oliveira (2018), os autores sugerem que entre os anos 2000 até 2013, houve uma tendência de aumento na padronização de modelo das companhias aéreas brasileiras. Por fim, o estudo ainda destaca que, havendo previsão de crescimento de mercado no curto prazo, as companhias aéreas são mais adeptas a reduzir a padronização de frota em função de receber mais rapidamente novas aeronaves, mesmo que provenientes do mercado secundário.

Outra pesquisa realizada sobre os efeitos da padronização sobre os custos operacionais e lucratividade de companhias aéreas brasileiras pode ser encontrada em Narcizo (2018). Neste estudo, modelos econométricos são utilizados para investigar, além dos efeitos sobre custos e lucros, a relação entre padronização e o modelo de negócio de baixo custo. A análise é baseada em regressões lineares, as quais trabalham com variáveis de padronização de frota e buscam entender como estas podem afetar o desempenho da empresa. Os dados utilizados na pesquisa compreendem quatro companhias aéreas brasileiras que operaram no período de janeiro de 2002 a dezembro de 2013, sendo elas: Avianca, Azul, Gol e Latam. Seguindo premissas da literatura existente, o estudo traz variáveis segmentadas em três grupos principais e analisa os impactos de cada uma delas no custo operacional e na margem de lucro. Quanto mais variáveis relevantes inseridas na equação, mais detalhada é a abordagem do problema. Entretanto, deve-se ter cuidado para que não sejam adicionadas variáveis não relacionadas ao problema e garantir um tratamento estatístico adequado para àquelas que forem inseridas, de forma a evitar a presença de viés na estimação dos resultados e garantir confiabilidade nas interpretações. Para os leitores mais interessados na econometria envolvida, recomenda-se a leitura de Wooldridge, 2013 e Gujarati, 2004. Os resultados do estudo estão descritos na Tabela 3, onde um sinal positivo sugere uma correlação positiva entre as variáveis e um sinal negativo sugere correlação negativa. Resultados expressos por "NS" correspondem a resultados com correlação estatisticamente não significante, sugerindo não haver relação estatística entre os termos analisados. Ressalta-se que, apesar da ausência de significância estatística, resultados NS também são relevantes, visto que permitem justamente afastar a hipótese de causalidade entre as variáveis.

Os fatores de custo têm por objetivo capturar o efeito que os gastos da empresa têm no custo operacional por assento-quilômetro e na margem de lucro operacional. Foram testados gastos com tripulantes, combustível, manutenção e arrendamento de aeronaves. As variáveis apresentaram um resultado bastante intuitivo. A relação positiva dos custos específicos com o custo operacional por assento-quilômetro segue a lógica de que, um aumento em qualquer um dos gastos, resultará em consequente aumento nos custos operacionais totais da companhia aérea. De forma antagônica, a relação negativa com margem de lucro operacional, sugere que além de aumentar o custo total, gastos maiores também reduzem a margem de lucro operacional. A exceção é observada nos gastos com arrendamento de aeronaves, o qual não apresentou a efeito. Uma das possibilidades para esse resultado é de que os custos com arrendamento são fixos e sofrem menor variação no curto prazo, aumentando a possibilidade de a companhia responder com aumento de tarifas para compensação e evita a redução na margem de lucro.

**Tabela 2 – Resumo dos resultados de Narcizo (2018)**

| Características | (1) Custo Operacional por Assento-Quilômetro | (2) Margem de Lucro Operacional |
|---|---|---|
| Custo com pilotos e comissários | + | - |
| Custo com combustível | + | - |
| Custo com manutenção | + | - |
| Custo com arrendamento | + | NS |
| RPK | NS | + |
| Número de destinos | NS | NS |
| Etapa de voo média | - | NS |
| Eficiência no consumo de combustível | - | NS |
| Eficiência operacional | NS | - |
| Tamanho de frota | NS | - |
| Idade média de frota | + | NS |
| Tamanho médio das aeronaves | - | NS |
| Padronização por modelo | NS | NS |
| Padronização por família | - | + |
| Padronização por fabricante | NS | NS |
| Padronização por família – LCC Gol | NS | NS |
| Padronização por família – LCC Azul | NS | NS |

*NS = efeito não significante.

Em relação aos fatores operacionais, o resultado obtido para a variável RPK indica efeito nulo sobre o custo operacional unitário. O resultado é contrário aos achados por Zou, Yu e Dresner (2015). Entretanto, a hipótese defendida neste estudo é de que, salvo se a companhia disponibilizar mais assentos, o custo de voo existente será muito próximo independentemente de a aeronave estar cheia ou vazia, não havendo, portanto, redução no custo unitário. Supõe-se que o efeito do RPK na forma como os autores propõem, seria mais bem observado se a variável dependente fosse custo por passageiro pagante-quilômetro (CRPK) e não custo por assento-quilômetro (CASK). Já os efeitos sobre a margem de lucro operacional são intuitivamente positivos, indicação de que companhias com maior quantidade de passageiros pagantes apresentam maior margem de lucro, como consequência de uma maior obtenção de receita.

Nenhum efeito do número de destinos foi encontrado sobre os custos operacionais unitários e sobre a margem de lucro operacional. Já etapa de voo média mostrou-se negativamente correlacionada com custos unitários, porém sem efeitos sobre a margem de lucro operacional, sugerindo que companhias com maior etapa média de voo até possuem menores custos unitários, porém não apresentam maior margem de lucro.

Assim como sugerido por Zou, Yu e Dresner (2015), a variável de eficiência de combustível apresentou efeito negativo nos custos operacionais, corroborando a hipótese de que companhias com maior quantidade de assentos-quilômetros voados por litro de querosene tendem a ter menores custos. Entretanto, a ausência de efeito sobre margem de lucro contradiz os resultados dos outros autores. A justificativa para tal resultado se dá na ideia de que a companhia pode optar por abaixar suas tarifas para manter-se mais competitiva e aumentar a sua participação no mercado, em vez de apresentar maior margem de lucro.

A variável de eficiência operacional não apresentou efeitos no custo unitário. A ideia é de que, por mais horas de voo que as aeronaves de uma companhia aérea possuam, apenas os custos fixos é que poderiam ser diluídos pela maior quantidade de assento-quilômetro oferecido. Gastos com combustível, manutenção, seguro e tripulação, os quais são os maiores das empresas brasileiras, também variariam conforme a quantidade de voos realizados, contrabalanceando a maior oferta de assentos. Curiosamente e apesar da ausência de efeitos nos custos, a eficiência impactou de forma negativa a margem de lucro. Neste caso, a explicação baseia-se na ideia de que quando a companhia coloca mais horas por aeronave, há um interesse da empresa em ter mais capital e aumentar sua fatia de mercado, induzindo a maiores gastos na tentativa de aumentar a eficiência de sua operação e de manter suas aeronaves em solo por menos tempo.

Tamanho de frota não apresentou influência sobre os custos operacionais unitários, sugerindo não haver economia de escala ao aumentar a frota de aeronaves. Por outro lado, nota-se o efeito negativo sobre margem de lucro operacional, o que indicaria um menor lucro para companhias com frotas muito grandes. De certa forma, o resultado pode ser interpretado no sentido de que a companhia precisaria ajustar sua frota ao tamanho do RPK para ter maiores lucros, já que, caso contrário, ela teria frota ociosa e uma consequente redução na sua margem operacional.

Os resultados para idade de frota foram opostos aos resultados obtidos por Merkert e Hensher (2011), Zuidberg (2014) e Zou, Yu e Dresner (2015). O efeito positivo em custos indica que frotas mais velhas apresentariam maiores gastos. Os autores haviam sugerido anteriormente que aeronaves mais novas apresentariam custos maiores, decorrentes de gastos elevados com arrendamento e depreciação. Entretanto, pode-se supor que no mercado brasileiro, o maior consumo de combustível e os maiores gastos com manutenção de aeronaves mais velhas têm maior peso nos custos operacionais unitários. Como última variável dos fatores operacionais, o tamanho médio das aeronaves apresentou impacto negativo sobre custos operacionais, sugerindo que aeronaves maiores apresentariam menores custos operacionais unitários, ideia similar à discutida por Merker e Hensher (2011).

Como fator central da pesquisa, as variáveis de padronização têm como premissa a ideia de que companhias aéreas com maior padronização possuiriam menores custos operacionais unitários. Segundo os resultados obtidos no estudo, apenas a padronização por família confirma tal premissa. Os resultados rejeitam tal hipótese quando considerando a padronização por fabricante e modelo de aeronave, indicando que tais níveis não afetam os custos operacionais unitários das companhias aéreas. Tais resultados são similares aos apresentados por Zou, Yu e Dresner (2015) e sugerem que não há benefícios em manter um único modelo de aeronave na frota ou em selecionar aeronaves de apenas um único fabricante.

De forma bastante similar, apenas a padronização de família apresentou efeito na margem de lucro operacional. O impacto positivo indica que a padronização de família tem potencial não somente para reduzir os custos operacionais, mas também de aumentar a margem de lucro das companhias aéreas, assim como os resultados sugeridos por Kilpi (2007), West e Bradley (2008), Brüggen e Klose (2010) e Zou, Yu e Dresner (2015).

A hipótese para tais resultados baseia-se na ideia de que modelos de aeronaves de uma mesma família possuem capacidade e tamanho diferentes, que permite a uma companhia ter uma maior flexibilidade e melhor captação de receita. Entretanto, como ainda assim são aeronaves similares, estas não perdem os benefícios da comunalidade

Notou-se ainda que o modelo de negócios de uma companhia aérea não tem efeito intensificador ou atenuador sobre as consequências da padronização de frota. Os resultados indicam que as variáveis de padronização de família com interação de companhias LCC não apresentaram efeitos, indicando que, independentemente do modelo de negócio adotado pela companhia aérea, a padronização de família terá o mesmo efeito nos custos unitários e na margem de lucro. Dessa forma, pode-se dizer que companhias aéreas com modelos de negócios classicamente classificados como full service ou low-cost (mesmo que híbridas) não tem maiores benefícios sobre a padronização.

## IV. Conclusões

Por qual o motivo então são altas as chances do meu voo ser operado com o mesmo modelo de aeronave? A realidade é que a resposta para essa pergunta pode variar por inúmeros motivos. Voar em uma mesma companhia em rotas similares pode fazer com que de fato o modelo de aeronave seja sempre igual, visto que dentre as aeronaves da empresa, esta provavelmente será a mais adequada para combinar a oferta e demanda daquele mercado. Companhias que operam em mercados e rotas diferentes pode fazer com que seja mais interessante possuir mais modelos de aeronaves, tal como comumente ocorre quando a empresa tem operação em rotas regionais ou internacionais de longo curso. Portanto, havendo mais de um modelo de aeronave na companhia em questão, voar sempre em um mesmo modelo de avião pode tratar-se apenas de uma coincidência.

Se, por outro lado, além da questão de adequar o modelo de aeronave à demanda do mercado, a companhia aérea possuir uma frota com poucos modelos ou com uma única família de aeronave, o ponto chave para responder à pergunta está na padronização. Como vimos, a escolha e seleção de aeronaves para composição da frota é um problema com muitas vertentes e estas devem ser analisadas de maneira conjunta. Ao optar por aeronaves que tenham elevada comunalidade, a empresa pode estar buscando qualquer um dos benefícios citados anteriormente, como a redução dos custos com tripulação e manutenção, a elevação na capacidade de recuperação de escala em caso de ruptura e aumento no poder de barganha com fabricantes.

Outro ponto bastante relevante para responder à pergunta principal do trabalho pode ser obtido a partir do estudo nacional mencionado. Os resultados encontrados em Narcizo (2018) sugerem uma maior concorrência na rota faz com que as companhias aéreas padronizem mais suas frotas em nível de família. Logo, passageiros corporativos, os quais costumeiramente se fidelizam mais, tendem a voar com a mesma empresa, e por isso, voarão em aeronaves do mesmo modelo com muita frequência devido à sua fidelização. Por outro lado, no segmento de lazer, companhias aéreas com maior padronização tendem a ter menores custos e, por decorrência, preços mais baixos. Logo, pode-se supor que a chance de voar com a mesma aeronave aumente, visto que ao buscarmos preços mais baixos, acabamos sem querer optando pelas companhias com maior padronização.

Quando bem ponderada, os benefícios de tornar-se fiel a uma família de aeronave pode superar muito os riscos associados a possuir apenas um conjunto de modelos na frota. Se, durante a crise com os modelos Boeing 737 MAX, a padronização de família da companhia aérea Gol foi um problema para a empresa suprir sua demanda, visto que sua única família de aeronaves estava parcialmente impossibilitada de voar, por outro lado, manter uma frota única trouxe simplicidade e reduziu os custos da operação durante a crise mundial gerada pelo vírus da COVID-19. Já a complexidade e custo maior decorrente da elevada diversidade de frota da Azul também se mostrou um grande trunfo durante a mesma crise sanitária. Por ser a única companhia aérea no país a operar aeronaves que vão desde 9 até mais de 200 assentos, a empresa dispõe de total flexibilidade para adaptar suas operações de forma a combinar oferta com demanda, apesar das condições de grande incerteza como as observadas no período. Tais exemplos mostram que a decisão de cada empresa sob sua frota é complexa e o ponto de equilíbrio entre adotar ou não a padronização de frota e que nível de padronização adotar será ditado pelo plano de negócios e pela estratégia estabelecida por cada companhia aérea.